# Fractions, Projective Representation, Duality, Linear Algebra and Geometry

## Vaclav Skala

*Department of Computer Science and Engineering, Faculty of Applied Sciences,
University of West Bohemia, Univerzitni 8, 306 14 Plzen, Czech Republic*
*http://www.VaclavSkala.eu*

**Abstract.** This contribution describes relationship between fractions, projective representation, duality, linear algebra and geometry. Many problems lead to a system of linear equations and this paper presents equivalence of the cross–product operation and solution of a system of linear equations $Ax = 0$ or $Ax = b$ using projective space representation and homogeneous coordinates. It leads to conclusion that division operation is not required for a solution of a system of linear equations, if the projective representation and homogeneous coordinates are used. An efficient solution on CPU and GPU based architectures is presented with an application to barycentric coordinates computation as well.

*Key words:* Linear system of equations, extended cross product, projective space computation, geometric algebra, scientific computation.

## 1 Introduction

Many applications, not only in computer vision, require a solution of a homogeneous system of linear equations $Ax = 0$ or a non-homogeneous system of linear equations $Ax = b$. However, the numerical solution actually does not allow further symbolic manipulation. Even more, solutions of equations $Ax = 0$ and $Ax = b$ are considered as different problems and especially $Ax = 0$ is not usually solved quite correctly as users tend to use some additional condition for $x$ unknown (usually setting $x_k = 1$ or so).

Many problems in computer vision, computer graphics and visualization are 3-dimensional. Therefore, specific numerical approaches can be applied to speed up the solution. In the following extended cross product, also called outer product or progressive product, is introduced in the "classical" notation using the " $\times$ " symbol.

## 2 Extended Cross-product

Let us consider the standard cross product of two vectors $a = [a_1, a_2, a_3]^T$ and $b = [b_1, b_2, b_3]^T$. Then the cross product is defined as:

$$a \times b = \det \begin{bmatrix} i & j & k \\ a_1 & a_2 & a_3 \\ b_1 & b_2 & b_3 \end{bmatrix} \quad (1)$$

where: $i = [1,0,0]^T, j = [0,1,0]^T, k = [0,0,1]^T$.

If a matrix form is needed, then we can write:

$$a \times b = \begin{bmatrix} 0 & -a_3 & a_2 \\ a_3 & 0 & -a_1 \\ -a_2 & a_1 & 0 \end{bmatrix} \begin{bmatrix} b_1 \\ b_2 \\ b_3 \end{bmatrix} \quad (2)$$

In some applications, the matrix form is more convenient.

Let us introduce the extended cross product of three vectors $a = [a_1, \ldots, a_n]^T$, $b = [b_1, \ldots, b_n]^T$ and $c = [c_1, \ldots, c_n]^T$, $n = 4$ as:

$$a \times b \times c = \det \begin{bmatrix} i & j & k & l \\ a_1 & a_2 & a_3 & a_4 \\ b_1 & b_2 & b_3 & b_4 \\ c_1 & c_2 & c_3 & c_4 \end{bmatrix} \quad (3)$$

where: $i = [1,0,0,0]^T$, $j = [0,1,0,0]^T$, $k = [0,0,1,0]^T$, $l = [0,0,0,1]^T$.

It can be shown that there exists a matrix form for the extended cross-product representation:

$$a \times b \times c = (-1)^{n+1} \begin{bmatrix} 0 & -\delta_{34} & \delta_{24} & -\delta_{23} \\ \delta_{34} & 0 & -\delta_{14} & \delta_{13} \\ -\delta_{24} & \delta_{14} & 0 & -\delta_{12} \\ \delta_{23} & -\delta_{13} & \delta_{12} & 0 \end{bmatrix} \begin{bmatrix} c_1 \\ c_2 \\ c_3 \\ c_4 \end{bmatrix} \quad (4)$$

where: $n = 4$. In this case and $\delta_{ij}$ are sub-determinants with columns $i, j$ of the matrix $T$ defined as:

$$T = \begin{bmatrix} a_1 & a_2 & a_3 & a_4 \\ b_1 & b_2 & b_3 & b_4 \end{bmatrix} \quad (5)$$

e.g. sub-determinant $\delta_{24} = \det \begin{bmatrix} a_2 & a_4 \\ b_2 & b_4 \end{bmatrix}$ etc.

The extended cross product for 5-dimensions is defined as:

$$a \times b \times c \times d = \det \begin{bmatrix} i & j & k & l & n \\ a_1 & a_2 & a_3 & a_4 & a_5 \\ b_1 & b_2 & b_3 & b_4 & b_5 \\ c_1 & c_2 & c_3 & c_4 & c_5 \\ d_1 & d_2 & d_3 & d_4 & d_5 \end{bmatrix} \quad (6)$$

where: $i = [1,0,0,0,0]^T$, $j = [0,1,0,0,0]^T$, $k = [0,0,1,0,0]^T$, $l = [0,0,0,1,0]^T$, $n = [0,0,0,0,1]^T$. It can be shown that there exists a matrix form as well:

$$a \times b \times c \times d$$
$$= (-1)^{n+1} \begin{bmatrix} 0 & -\delta_{345} & \delta_{245} & -\delta_{235} & \delta_{234} \\ \delta_{345} & 0 & -\delta_{145} & \delta_{135} & -\delta_{134} \\ -\delta_{245} & \delta_{145} & 0 & -\delta_{125} & \delta_{124} \\ \delta_{235} & -\delta_{135} & \delta_{125} & 0 & -\delta_{123} \\ -\delta_{234} & \delta_{134} & -\delta_{124} & \delta_{123} & 0 \end{bmatrix} \begin{bmatrix} d_1 \\ d_2 \\ d_3 \\ d_4 \\ d_5 \end{bmatrix} \quad (7)$$

where $n = 5$. In this case and $\delta_{ijk}$ are sub-determinants with columns $i, j, k$ of the matrix $T$ defined as:

$$T = \begin{bmatrix} a_1 & a_2 & a_3 & a_4 & a_5 \\ b_1 & b_2 & b_3 & b_4 & b_5 \\ c_1 & c_2 & c_3 & c_4 & c_5 \end{bmatrix} \quad (8)$$

e.g. sub-determinant $\delta_{245}$ is defined as:

$$\delta_{245} = \det \begin{bmatrix} a_2 & a_4 & a_5 \\ b_2 & b_4 & b_5 \\ c_2 & c_4 & c_5 \end{bmatrix}$$
$$= a_2 \det \begin{bmatrix} b_4 & b_5 \\ c_4 & c_5 \end{bmatrix} - a_4 \det \begin{bmatrix} b_2 & b_5 \\ c_2 & c_5 \end{bmatrix} + a_5 \det \begin{bmatrix} b_2 & b_4 \\ c_2 & c_4 \end{bmatrix} \quad (9)$$

In spite of the "complicated" description above, this approach leads to a faster computation in the case of lower dimensions [1],[2].

## 3   Extended Cross-product

Projective representation and its application for computation is considered mysterious or too complex. Nevertheless, we are using it naturally very frequently in the form of fractions, e.g. $a/b$. We also know that fractions help us to express values, which cannot be expressed precisely due to limited length of a mantissa, e.g. $1/3 = 0{,}33\ldots\ldots.333\ldots = 0.\overline{3}$. In the following, we will explore projective representation, actually rational fractions, and its applicability.

### 3.1   Projective representation

Projective extension of the Euclidean space is used commonly in computer graphics and computer vision mostly for geometric transformations. However, in computational sciences, the projective representation is not used, in general. This chapter shortly introduces basic properties and mutual conversions.

The given point $X = (X, Y)$ in the Euclidean space $E^2$ is represented in homogeneous coordinates as $x = [x, y: w]^T$, $w \neq 0$. It can be seen that $x$ is actually a line in the projective space $P^3$ with the origin excluded. Mutual conversions are defined as:

$$X = \frac{x}{w} \qquad\qquad Y = \frac{y}{w} \quad (10)$$

where: $w \neq 0$ is the homogeneous coordinate. Note that the homogeneous coordinate $w$ is actually a scaling factor with no physical meaning, while $x, y$ are values with physical units in general.

The projective representation enables us nearly double precision as the mantissa of $x$, resp. $y$ and $w$ are used for a value representation. However, we have to distinguish two different data types, i.e.

- Projective representation of a $n$-dimensional value $X = (X_1, \ldots, X_n)$, represented by one-dimensional array $x = [x_1, \ldots, x_n : x_w]^T$, e.g. coordinates of a point, that is fixed to the origin.
- Projective representation of a $n$-dimensional vector (in the mathematical meaning) $A = (A_1, \ldots, A_n)$, represented by one dimensional array $a = [a_1, \ldots, a_n : a_w]^T$. In this case, the homogeneous coordinate $a_w$ is actually just a scaling factor. Any vector is not fixed to the origin of the coordinate system and it is "movable".

Therefore, a user should take an attention to the correctness of operations.

## 3.2 Principle of Duality

The projective representation offers also one very important property – principle of duality. The principle of duality in $E^2$ states that any theorem remains true when we interchange the words "point" and "line", "lie on" and "pass through", "join" and "intersection", "collinear" and "concurrent" and so on. Once the theorem has been established, the dual theorem is obtained as described above In other words, the principle of duality says that in all theorems it is possible to substitute the term "point" by the term "line" and the term "line" by the term "point" etc. in $E^2$ and the given theorem stays valid. Similar duality is valid for $E^3$ as well, i.e. the terms "point" and "plane" are dual etc. it can be shown that operations "join" a "meet" are dual as well.

This helps a lot to solve some geometrical problems. In the following, we will demonstrate that on very simple geometrical problems like intersection of two lines, resp. three planes and computation of a line given by two points, resp. of a plane given by three points.

## 4  Solution of $Ax = b$

Solution of non-homogeneous system of equation $AX = b$ is used in many computational tasks. For simplicity of explanation, let us consider a simple example of intersection computation of two lines $p_1$ and $p_2$ in $E^2$ given as:
$$p_1: A_1 X + B_1 Y + C_1 = 0 \qquad p_2: A_2 X + B_2 Y + C_2 = 0 \qquad (11)$$
An intersection point of two those lines is given as a solution of a linear system of equations: $Ax = b$:
$$\begin{bmatrix} a_1 & b_1 \\ a_2 & b_2 \end{bmatrix} \begin{bmatrix} X \\ Y \end{bmatrix} = \begin{bmatrix} -c_1 \\ -c_2 \end{bmatrix} \qquad (12)$$
Generally, for the given system of $n$ liner equations with $n$ unknowns in the form $AX = b$ the solution is given:
$$X_i = \frac{\det(A_i)}{\det(A)} \qquad i = 1, \dots, n \qquad (13)$$
where: $A$ is a regular matrix $n \times n$ having non-zero determinant, the matrix $A_i$ is the matrix $A$ with replaced $i^{th}$ column by the vector $b$ and $X = [X_1, \dots, X_n]^T$ is a vector of unknown values.

In a low dimensional case using general methods for solution of linear equations, e.g. Gauss-Seidel elimination etc., is computational expensive. In addition, division operation is computationally expensive and decreasing precision of a solution.

Usually, a condition **if** $\det(A) < eps$ **then** EXIT is taken for solving "close to singular cases". Of course, nobody knows, what a value of $eps$ is appropriate.

## 5 Solution of $Ax = 0$

There is another very simple geometrical problem; determination of a line $p$ given by two points $X_1 = (X_1, Y_1)$ and $X_2 = (X_2, Y_2)$ in $E^2$. This seems to be a quite simple problem as we can write:
$$aX_1 + bY_1 + c = 0 \qquad aX_2 + bY_2 + c = 0 \qquad (14)$$
i.e. it leads to a solution of homogeneous systems of equations $AX = 0$, i.e.:
$$\begin{bmatrix} X_1 & Y_1 & 1 \\ X_2 & Y_2 & 1 \end{bmatrix} \begin{bmatrix} a \\ b \\ c \end{bmatrix} = 0 \qquad (15)$$
In this case, we obtain one parametric set of solutions as the Eq.(15) can be multiplied by any value $q \neq 0$ and the line is the same.

There is a problem – we know that lines and points are dual in the $E^2$ case, so the question is why the solutions are not dual. However, if the projective representation is used the duality principle will be valid, as follows.

## 6 Solutions of $Ax = b$ and $Ax = 0$

Let us consider again intersection of two lines $p_1 = [a_1, b_1 : c_1]^T$ and $p_2 = [a_2, b_2 : c_2]^T$ leading to a solution of non-homogeneous linear system $AX = b$, which is given as:
$$p_1 : a_1 X + b_1 Y + c_1 = 0 \qquad p_2 : a_2 X + b_2 Y + c_2 = 0 \qquad (16)$$
If the equations are multiplied by $w \neq 0$ we obtain:
$$\begin{matrix} p_1 : a_1 X + b_1 Y + c_1 \triangleq & p_2 : a_2 X + b_2 Y + c_2 \triangleq \\ a_1 x + b_1 y + c_1 w = 0 & a_2 x + b_2 y + c_2 w = 0 \end{matrix} \qquad (17)$$
where: $\triangleq$ means „projectively equaivalent to" as $x = wX$ and $y = wY$.

Now we can rewrite the equations to the matrix form as $Ax = 0$:
$$\begin{bmatrix} a_1 & b_1 & -c_1 \\ a_2 & b_2 & -b_2 \end{bmatrix} \begin{bmatrix} x \\ y \\ w \end{bmatrix} = \begin{bmatrix} 0 \\ 0 \end{bmatrix} \qquad (18)$$
where $x = [x, y : w]^T$ is the intersection point in the homogeneous coordinates.

In the case of computation of a line given by two points given in homogeneous coordinates, i.e. $x_1 = [x_1, y_1 : w_1]^T$ and $x_2 = [x_2, y_2 : w_2]^T$, the Eq.(14) is multiplied by $w_i \neq 0$. Then, we get a solution in the matrix form as $Ax = 0$, i.e.
$$\begin{bmatrix} x_1 & y_1 & w_1 \\ x_2 & y_2 & w_2 \end{bmatrix} \begin{bmatrix} a \\ b \\ c \end{bmatrix} = 0 \qquad (19)$$
Now, we can see that the formulation is leading in the both cases to the same numerical problem: to a solution of a homogeneous linear system of equations.

However, a solution of homogeneous linear system of equations is not quite straightforward as there is a one parametric set of solutions and all of them are projectively equivalent. It can be seen that the solution of Eq. (18), i.e. intersection of two lines in $E^2$, is equivalent to:
$$x = p_1 \times p_2 \qquad (20)$$

Due to the principle of duality we can write for a line given by two points:
$$\boldsymbol{p} = \boldsymbol{x}_1 \times \boldsymbol{x}_2 \quad (21)$$
In the three dimensional case we can use extended cross product.

A plane $\rho: aX + bY + cY + d = 0$ given by points $\boldsymbol{x}_1 = [x_1, y_1, z_1: w_1]^T$, $\boldsymbol{x}_2 = [x_2, y_2, z_2: w_2]^T$ and $\boldsymbol{x}_2 = [x_3, y_3, z_3: w_3]^T$ is determined in the projective representation as:
$$\boldsymbol{\rho} = [a, b, c: d]^T = \boldsymbol{x}_1 \times \boldsymbol{x}_2 \times \boldsymbol{x}_2 \quad (22)$$
and the intersection point $\boldsymbol{x}$ of three planes points $\boldsymbol{\rho}_1 = [a_1, b_1, c_1: d_1]^T$, $\boldsymbol{\rho}_2 = [a_2, b_2, c_2: d_2]^T$ and $\boldsymbol{\rho}_3 = [a_3, b_3, c_3: d_3]^T$ is determined in the projective representation as:
$$\boldsymbol{x} = [x, y, z: w]^T = \boldsymbol{\rho}_1 \times \boldsymbol{\rho}_2 \times \boldsymbol{\rho}_2 \quad (23)$$
due to the duality principle.

It can be seen that there is no division operation needed, if the result can be left in the projective representation.

The approach presented above has another one great advantage as it allows symbolic manipulation as we have avoided numerical solution and precision is nearly doubled.

## 7  Barycentric coordinates

The barycentric coordinates are often used in many engineering applications, not only in geometry. The barycentric coordinates computation leads to a solution of a system of linear equations. However it was shown, that a solution of a linear system equations is equivalent to the extended cross product [3],[4],[5]. Therefore, it is possible to compute the barycentric coordinates using cross product, which is convenient for application of SSE instructions or for GPU oriented computations. Let us demonstrate the proposed approach on a simple example again.

Given a triangle in $E^2$ defined by points $\boldsymbol{x}_i = [x_i, y_i: 1]^T$, $i = 1, \ldots, 3$, the barycentric coordinates of the point $\boldsymbol{x}_0 = [x_0, y_0: 1]^T$ can be computed as follows:
$$\begin{aligned} \lambda_1 x_1 + \lambda_2 x_2 + \lambda_3 x_3 &= x_0 \\ \lambda_1 y_1 + \lambda_2 y_2 + \lambda_3 y_3 &= y_0 \\ \lambda_1 + \lambda_2 + \lambda_3 &= 1 \end{aligned} \quad (24)$$
For simplicity, we set $w_i = 1, i = 1, \ldots, 3$. It means that we have to solve a system of linear equations $\boldsymbol{Ax} = \boldsymbol{b}$:
$$\begin{bmatrix} x_1 & x_2 & x_3 \\ y_1 & y_2 & y_3 \\ 1 & 1 & 1 \end{bmatrix} \begin{bmatrix} \lambda_1 \\ \lambda_2 \\ \lambda_3 \end{bmatrix} = \begin{bmatrix} x_0 \\ y_0 \\ 1 \end{bmatrix} \quad (25)$$
if the points are given in the projective space with homogeneous coordinates $\boldsymbol{x}_i = [x_i, y_i: w_i]^T$, $i = 1, \ldots, 3$ and $\boldsymbol{x}_0 = [x_0, y_0: w_0]^T$. It can be proved, due to the multilinearity, we need to solve a linear system $\boldsymbol{Ax} = \boldsymbol{b}$:

$$\begin{bmatrix} x_1 & x_2 & x_3 \\ y_1 & y_2 & y_3 \\ w_1 & w_2 & w_3 \end{bmatrix} \begin{bmatrix} \lambda_1 \\ \lambda_2 \\ \lambda_3 \end{bmatrix} = \begin{bmatrix} x_0 \\ y_0 \\ w_0 \end{bmatrix} \quad (26)$$

Let us define new vectors containing a row of the matrix $A$ and vector $b$ as:
$$\boldsymbol{x} = [x_1, x_2, x_3, x_0]^T \quad \boldsymbol{y} = [y_1, y_2, y_3, y_0]^T \quad \boldsymbol{w} = [w_1, w_2, w_3, w_0]^T \quad (27)$$

The projective barycentric coordinates $\boldsymbol{\xi} = [\xi_1, \xi_2, \xi_3 : \xi_w]^T$ are given as:
$$\lambda_1 = -\frac{\xi_1}{\xi_w} \quad \lambda_2 = -\frac{\xi_2}{\xi_w} \quad \lambda_3 = -\frac{\xi_3}{\xi_w} \quad (28)$$

i.e.
$$\lambda_i = -\frac{\xi_i}{\xi_w} \qquad i = 1, \ldots, 3 \quad (29)$$

Using the extended cross product, the projective barycentric coordinates are given as:
$$\boldsymbol{\xi} = \boldsymbol{x} \times \boldsymbol{y} \times \boldsymbol{w} = \det \begin{bmatrix} \boldsymbol{i} & \boldsymbol{j} & \boldsymbol{k} & \boldsymbol{l} \\ x_1 & x_2 & x_3 & x_0 \\ y_1 & y_2 & y_3 & y_0 \\ w_1 & w_2 & w_3 & w_4 \end{bmatrix} \quad (30)$$
$$= [\xi_1, \xi_2, \xi_3 : \xi_w]^T$$

where $\boldsymbol{i} = [1,0,0,0]^T, \boldsymbol{j} = [0,1,0,0]^T, \boldsymbol{k} = [0,0,1,0]^T, \boldsymbol{l} = [0,0,0,1]^T$.

Similarly in the $E^3$ case, given a tetrahedron in $E^3$ defined by points $\boldsymbol{x}_i = [x_i, y_i, z_i : w_i]^T$, $i = 1, \ldots, 3$, and the point $\boldsymbol{x}_0 = [x_0, y_0, z_0 : w_0]^T$:
$$\boldsymbol{x} = [x_1, x_2, x_3, x_4 : x_0]^T \qquad \boldsymbol{y} = [y_1, y_2, y_3, y_4 : y_0]^T$$
$$\boldsymbol{z} = [z_1, z_2, z_3, z_4 : z_0]^T \qquad \boldsymbol{w} = [w_1, w_2, w_3, w_4 : w_0]^T \quad (31)$$

Then the projective barycentric coordinates are given as:
$$\boldsymbol{\xi} = \boldsymbol{x} \times \boldsymbol{y} \times \boldsymbol{z} \times \boldsymbol{w} = [\xi_1, \xi_2, \xi_3, \xi_4 : \xi_w]^T \quad (32)$$

The Euclidean barycentric coordinates are given as:
$$\lambda_1 = -\frac{\xi_1}{\xi_w} \quad \lambda_2 = -\frac{\xi_2}{\xi_w} \quad \lambda_3 = -\frac{\xi_3}{\xi_w} \quad \lambda_4 = -\frac{\xi_4}{\xi_w} \quad (33)$$

i.e.
$$\lambda_i = -\frac{\xi_i}{\xi_w} \qquad i = 1, \ldots, 4 \quad (34)$$

It can be seen that method of the barycentric coordinates is simple and convenient for vector-vector operations especially if SSE instructions or GPU is used.

## 8 GPU implementation

Many today's computational systems can use GPU support, which allows fast and parallel processing. The above presented approach offers significant speed up as the "standard" cross product is implemented in hardware as an instruction and the extended cross-product for 4D can be implemented as:

```
float4 cross_4D(float4 x1, float4 x2, float4 x3)
{float4 a;
  a.x = dot(x1.yzw, cross(x2.yzw, x3.yzw));
  a.y = -dot(x1.xzw, cross(x2.xzw, x3.xzw));
  a.z = dot(x1.xyw, cross(x2.xyw, x3.xyw));
  a.w = -dot(x1.xyz, cross(x2.xyz, x3.xyz));
  return a}
```

In general, it can be seen that a solution of linear systems of equations on GPU for a small dimension $n$ is simple, fast and can be performed in parallel.

## 9 Conclusion

Projective representation is not widely used for general computation as it is mostly considered as applicable to computer graphics and computer vision field only. In this paper, the equivalence of cross product and solution of linear system of equations has been presented. The presented approach is especially convenient for 3-dimensional and 4 dimensional cases applicable in many engineering and statistical computations, in which significant speed up can be obtained using SSE instructions or GPU use. In addition, the presented approach enables symbolic manipulation as the solution of a system of linear equations is transformed to extended cross product using a matrix form, which enables symbolic manipulations.

Direct application of the presented approach has also been demonstrated on the barycentric coordinates computation and simple geometric problems.

The presented approach enables avoiding division operations, as a denominator is actually stored in the homogeneous coordinate w. It that leads to significant computational savings, increase of precision and robustness as the division operation is the longest one and the most decreasing precision of computation.

The above presented approach is based on author's recent publications mentioned in References below.

## Acknowledgements

The author would like to thank to colleagues at the University of West Bohemia in Plzen for discussions and to anonymous reviewers for their comments and hints, which helped to improve the manuscript significantly.
Research was supported by the MSMT CZ, project No.LH12181.

# References

[1] Skala,V.: Plücker Coordinates and Extended Cross Product for Robust and Fast Intersection Computation, CGI 2016 proccedings, ACM, pp.57-60, Greece, 2016

[2] Skala,V.: "Extended Cross-product" and Solution of a Linear System of Equations, Computational Science and Its Applications, ICCSA 2016, LNCS 9786, Vol.I, pp.18-35, Springer, 2016

[3] Skala,V.: Barycentric Coordinates Computation in Homogeneous Coordinates, Computers & Graphics, Elsevier, ISSN 0097-8493, Vol. 32, No.1, pp.120-127, 2008

[4] Skala,V.: Intersection Computation in Projective Space using Homogeneous Coordinates, Int. Journal of Image and Graphics, Vol.7, No.4, pp.615-628, 2008

[5] Skala,V.: Length, Area and Volume Computation in Homogeneous Coordinates, Int. Journal of Image and Graphics, Vol.6., No.4, pp.625-639, 2006

[6] Skala,V.: GPU Computation in Projective Space Using Homogeneous Coordinates , Game Programming GEMS 6 (Ed.Dickheiser,M.), pp.137-147, Charles River Media, 2006

[7] Skala,V.: A new approach to line and line segment clipping in homogeneous coordinates, The Visual Computer, Vol.21, No.11, pp.905-914, Springer Verlag, 2005